\documentstyle[12pt]{article}

\newcommand{\be}{\begin{equation}}
\newcommand{\ee}{\end{equation}}
\newcommand{\ba}{\begin{eqnarray}}
\newcommand{\ea}{\end{eqnarray}}
\newcommand{\lb}{\label}

\newcommand{\ds}{\displaystyle}
\newcommand{\ra}{\rightarrow}
\newcommand{\ol}{\overline}
\newcommand{\st}{^{\ast}}
\newcommand{\bb}[1]{\bibitem{#1}}
\begin{document}
\begin{titlepage}
\setcounter{page}{1}
\title{Gravitating Chern-Simons vortices}
\author{G\'erard Cl\'ement\thanks{E-mail:
 GECL@CCR.JUSSIEU.FR.} \\
\small Laboratoire de Gravitation et Cosmologie Relativistes
 \\
\small Universit\'e Pierre et Marie Curie, CNRS/URA769 \\
\small Tour 22-12, Bo\^{\i}te 142 \\
\small 4, place Jussieu, 75252 Paris cedex 05, France}
\bigskip
\date{\small December 11, 1995}
\maketitle
\begin{abstract}
The construction of self-dual vortex solutions to the Chern-Simons-Higgs
model (with a suitable eighth-order potential) coupled to Einstein gravity
in (2 + 1) dimensions is reconsidered. We show that the self-duality
condition may be derived from the sole assumption $g_{00} = 1$. Next, we
derive a family of exact, doubly self-dual vortex solutions, which
interpolate between the symmetrical and asymmetrical vacua. The
corresponding spacetimes have two regions at spatial infinity. The
eighth-order Higgs potential is positive definite, and closed timelike
curves are absent, if the gravitational constant is chosen to be negative.
\end{abstract}
\end{titlepage}

The well-known Abelian Higgs model \cite{N+O} can be generalized,
when the number of dimensions of Minkowski spacetime is restricted to three,
by the addition of a Chern-Simons term, leading to a model which has been
shown to admit charged vortex solutions \cite{MCSH}. While these are very
complicated, it was shown \cite{CSH} that if the Maxwell term is omitted
from the action\footnote{Because the Chern-Simons term dominates the
Maxwell term at long distances, this may be viewed as an asymptotic
approximation.}, and a specific sixth-order potential is chosen, the
theory admits self-dual vortex solutions.

The Maxwell-Chern-Simons-Higgs model may again be generalized by coupling it
to Einstein gravity. One expects \cite{Linet90} that this gravitating
model still admits vortex solutions with a spacetime metric asymptotic to
that generated, in (2 + 1)- dimensional general relativity, by a massive
spinning particle \cite{DJH}. It was shown in \cite{Valt} that the
Einstein-Chern-Simons-Higgs (ECSH) model (without the Maxwell term) admits
self-dual stationary solutions corresponding to systems of non-interacting
vortices, provided a suitable eighth-order potential is chosen. In \cite{C+L}
an Einstein-Maxwell-Chern-Simons-Higgs model with an additional real scalar
field was studied, and self-dual vortex solutions were similarly obtained
for a suitably chosen potential. When a certain limit in the space of
model parameters was taken, this model reduced to the ECSH model, and the
multi-vortex solutions of \cite{Valt} were recovered, with some generalization.

In a recent paper \cite{Lon}, the ECSH model was
reinvestigated, and it was shown that, under the same assumptions as in
\cite{Valt}, complemented by the assumption of rotational symmetry, the
full set of field equations (including the Einstein equations) can be
reduced to a set of four first-order differential equations. This does
not, however, represent a significant advance over the results of
\cite{Valt} \cite{C+L}, where the solution of the multi-vortex problem was
implicitly reduced to that of two (second-order) non-linear Schr\"odinger
equations (see below).

The purpose of this Letter is two-fold. First, the self-dual vortex
solutions of [6-8] were obtained under several assumptions, including the
condition $g_{00} = 1$, and the self-duality condition, eq. (\ref{S3}) below.
The fact that these {\em a priori} unrelated assumptions turn out to be
consistent leads us to surmise that one can be deduced from the other.
Indeed, we shall show in the following that the sole ansatz of a
stationary metric with $g_{00} = 1$ is enough to lead to the self-dual
solutions of \cite{Valt} \cite{C+L}, which we shall derive in a more
transparent, deductive fashion.

Our second motivation is the search for exact vortex solutions. As we have
mentioned, the field equations of the ECSH model may be, under the
assumption $g_{00} = 1$, partially integrated to a fourth-order
differential system, which at first sight can only be solved numerically.
However we shall show that a certain ``double self-duality'' ansatz yields
exact ``extreme'' vortex solutions. A peculiarity of these solutions is
their non-Euclidean spatial topology --- spatial sections have two
disconnected regions at infinity.

The abelian ECSH model is defined by the action
\ba \lb{action}
& I = {\ds \int} d^3x & \!\! \{\, -{\ds \frac{1}{16\pi G}}\,\sqrt{|g|} R
- {\ds \frac{\mu}{4}}\,\varepsilon^{\mu \nu \rho}F_{\mu\nu}A_{\rho}\nonumber \\
&& + \sqrt{|g|}\,[\, g^{\mu \nu}D_{\mu}\phi\st D_{\nu}\phi
- V(\phi, \phi\st )\,]\,\}\,,
\ea
where $F_{\mu \nu} = \partial_{\mu}A_{\nu} - \partial_{\nu}A_{\mu}$,
$D_{\mu}\phi = (\partial_{\mu} + ie A_{\mu})\phi$, and $\varepsilon^{\mu
\nu \rho}$ is the antisymmetric symbol; the Higgs potential $V(\phi,
\phi\st)$ shall be specified later on. The resulting Euler-Lagrange
equations are
\ba \lb{eqs}
&& R_{\mu \nu} = 8\pi G\, [ D_{\mu}\phi\st D_{\nu}\phi +
D_{\nu}\phi\st D_{\mu}\phi - 2Vg_{\mu \nu}]\,, \nonumber \\
&& \frac{\mu}{2\sqrt{|g|}}\varepsilon^{\mu \lambda \nu}F_{\mu \lambda} = i
e\,g^{\nu \rho}\,(\phi\st D_{\rho}\phi - D_{\rho}\phi\st \phi)\,,\\
&& \frac{1}{\sqrt{|g|}}\,D_{\mu}(\sqrt{|g|}\,g^{\mu \nu}D_{\nu}\phi) = -
\frac{\partial V}{\partial \phi\st }\,. \nonumber
\ea
We make the ansatz for stationary solutions
\ba \lb{stat}
& ds^2 & \equiv N^2\,(\,dt + \omega_i\,dx^i)^2 + \gamma_{ij}\,dx^i\,dx^j\,,
\nonumber\\
& A_{\mu}\,dx^{\mu} & \equiv A_0\,(\,dt + \omega_i\,dx^i) +
\ol{A}_i\,dx^i\,,
\ea
where the fields $N$, $A_0$, $\omega_i$, $\ol A_i$ and $\gamma_{ij}$ depend
only on the spatial variables $x^i$ ($i = 1,2$), and the spatial metric
$\gamma_{ij}$ will be used to move indices. We shall use in the following
the decomposition of the Ricci tensor into 2-scalar, 2-vector and 2-tensor
components,
\ba \lb{Ricci}
&& R_{00} \equiv N^2\,[\, -N^{-1}N^{;k}_{\;\;;k} + \frac{1}{2}\,b^2 \,]\,,
\nonumber \\
&& R_0^{\;i} \equiv \frac{1}{2
\sqrt{|\gamma|}}\,N^{-1}\varepsilon^{ij}(N^2b)_{;j}\,, \\
&&R^{ij} \equiv -N^{-1}N^{;i;j} + \frac{1}{2}\,[R_{\ds \gamma} - b^2]\;
\gamma^{ij}\,, \nonumber
\ea
where
\be \lb{twist}
b \equiv \frac{\;N}{\sqrt{|\gamma|}}\,\varepsilon^{ij}\omega_{j;i}\,,
\ee
is the ``twist'' field associated with the metric (\ref{stat}), and $;j$
is the spatial covariant derivative, as well as the reduced magnetic field
\be \lb{magn}
\ol{B} \equiv \frac{\;N}{\sqrt{|\gamma|}}\,\varepsilon^{ij}\ol{A}_{j;i}\,.
\ee

In order for static multi-vortex configurations to extremize the action
(\ref{action}), a balance must be achieved between the various forces
acting on the system. In the case $G = 0$ --- vortices in Minkowski
spacetime --- the conditions for such a balance are the Bogomol'nyi
\cite{Bogo} self-duality conditions. We do not expect this static equilibrium
to be altered (at least in lowest order) when the gravitational coupling is
switched on, provided the static gravitational force acting on the system
vanishes, $g_{00}(x^i) =$ constant. The inference is that this last condition,
which may be written (up to a rescaling of time)
\be \lb{ans1}
N = 1\,,
\ee
would appear to be necessary for (static) multi-vortex solutions to exist.
We now show that the condition (\ref{ans1}) is also sufficient, and
implies the self-duality conditions, if the potential has the appropriate
eighth-order form.

The ansatz (\ref{ans1}) implies, from the last equation (\ref{Ricci}) and the
Einstein field equations (\ref{eqs}), the proportionality relation
\be \lb{S1}
D^i\phi\st D^j\phi + D^j\phi\st D^i\phi \;\propto\; \gamma^{ij}\,,
\ee
where the contravariant components $D^i\phi$ are related to the reduced
covariant derivatives of the Higgs field by $D^i\phi =
\gamma^{ij}\,\ol{D}_j\phi$. Choosing for convenience conformal spatial
coordinates such that
\be \lb{conf}
\gamma_{ij} = - {\rm e}^{2u}\,\delta_{ij}\,,
\ee
we can write the condition (\ref{S1}) as
\be \lb{S2}
\ol{D}_z\phi\,\ol{D}_z\phi\st  = 0
\ee
(with $z \equiv x + iy$), which is solved either by $\ol{D}_z \phi = 0$ or
by $\ol{D}_z \phi\st = 0$. Changing back to an arbitrary spatial
coordinate system, we thus find that the solution of equation (\ref{S1}) is
given by the covariant self-duality condition \cite{C+L}
\be \lb{S3}
\ol{D}_i\phi = \mp \frac{i}{\sqrt{|\gamma|}}\,
\gamma_{ij}\varepsilon^{jk}\,\ol{D}_k\phi\,.
\ee
Using once this condition, we then find that the current on the right-hand
side of the second equation (\ref{eqs}) (the field-current relation) for
$\nu = i$ is a gradient, so that this equation can be integrated to
\be \lb{A_0}
A_0 = \pm \frac{e}{\mu}(\phi\st \phi - \eta^2)\,,
\ee
where $\eta^2$ is an arbitrary real constant. The Einstein equation
(\ref{eqs}) for the mixed component $R_0^{\;i}$ similarly integrates to
\be \lb{b}
b = -8\pi G\mu\,(A_0^2 - a^2)\,,
\ee
where $a^2$ is a new real constant\footnote{This constant, $-C$ in
\cite{C+L}, is absent in \cite{Valt}.}. The second equation (\ref{eqs})
for $\nu = 0$ then gives the reduced magnetic field in terms of $A_0$ and $b$:
\be \lb{Bbar}
\ol{B} = - A_0\,[\,\frac{2e^2}{\mu}\phi\st \phi + b\,]\,.
\ee

We are now in a position to exhibit the potential for which our ansatz
(\ref{ans1}) is indeed a solution of the field equations. The consistency
condition for this ansatz is a constraint for the component $R_{00}$ of
the Ricci tensor which, on account of equations (\ref{A_0}) and (\ref{b}),
yields the potential:
\be \lb{V}
V = \frac{e^4}{\mu^2}\phi\st \phi\,(\phi\st \phi - \eta^2)^2 - 2\pi G
\mu^2\,[\, \frac{e^2}{\mu^2}(\phi\st \phi - \eta^2)^2 - a^2 \,]^2\,.
\ee
Only two of the field equations (\ref{eqs}) remain. The Higgs equation is
redundant, by virtue of the Bianchi identities, while the contracted
Einstein equation for $R^{ij}$ yields the spatial two-curvature
\be \lb{gamma}
R_{\ds \gamma} = b^2 + 16\pi G\,[\,\gamma^{ij} \ol{D}_i\phi\st \ol{D}_j\phi
- 2V\,]\,.
\ee

Finally, we briefly summarize the integration of these equations along the
lines of \cite{C+L}. The self-duality conditions (\ref{S3}) can be solved
for the $\ol{A}_i$, leading to the expression of the reduced magnetic
field, written using the conformal metric (\ref{conf}),
\be \lb{Bbar2}
\ol{B} = \mp \frac{1}{e}\,{\rm e}^{-2u}\nabla^2 \ln{|\phi|}\,.
\ee
We may also choose a time gauge in which $\partial_i \omega_i = 0$ in
conformal coordinates, so that $\omega_i$ is the curl of a real potential
$\omega$, leading to the twist
\be \lb{b2}
b = - {\rm e}^{-2u} \nabla^2 \omega\,.
\ee
Then, the Higgs equation (\ref{eqs}) may be used to express the right-hand
side of (\ref{gamma}) as a linear combination of $\Box (\phi\st \phi)$,
$\ol{B}$ and $b$. These last two fields being covariant laplacians, as
well as $R_{\ds \gamma} = 2{\rm e}^{-2u} \nabla^2 u$ in conformal
coordinates, equation (\ref{gamma}) may be integrated to give the metric
function
\be \lb{h}
\sqrt{|\gamma|} = {\rm e}^{2u} = |h(z)|^{-2}\,|\phi|^{\ds 16 \pi G
\eta^2} {\rm e}^{\ds -8 \pi G (\,|\phi|^2 + 2 \mu a^2 \omega\,)}
\ee
(the arbitrariness of the analytical function $h(z)$ reflects the
invariance of the parametrization (\ref{conf}) under conformal
transformations). The problem is then reduced to the solution of two
coupled non-linear Schr\"odinger equations for the real functions $|\phi|$
and $\omega$, obtained by identifying (\ref{Bbar2}) to (\ref{Bbar}) and
(\ref{b2}) to (\ref{b}), respectively.

For solutions to these equations to qualify as vortex solutions, they
should obey appropriate boundary conditions. The conditions chosen in
\cite{C+L}
\be \lb{bc}
b(\infty) = 0, \;\; \ol{B}(\infty) = 0 \,,
\ee
ensure the convergence of the integrals giving the net spin and
magnetic flux in an asymptotically flat space (and asymptotic flatness
itself if $\Box (\phi\st \phi)$ goes to zero at infinity). These two
conditions are solved either by
\be \lb{bc1}
|\phi|(\infty) = \eta\,, \;\;A_0(\infty) = 0 \;\;\;{\rm with}\; a = 0\,,
\ee
or by
\be \lb{bc2}
|\phi|(\infty) = 0\,, \;\;A_0(\infty) = \mp a = \mp \frac{e}{\mu}\,
\eta^2\,.
\ee
If the spatial topology is $R^2$, the first set of boundary conditions
leads to topological solitons, while the second set leads to
non-topological solitons. In the following, we shall generalize these
conditions to allow for the possibility of non-asymptotically flat spatial
sections.

To find exact vortex solutions, we assume that the gauge field is also
self-dual in the sense of electric-magnetic duality \cite{K+K}, $E =
\varepsilon B$ (with $\varepsilon^2 = 1$). This may be enforced
covariantly by assuming that the scalar $F_{\mu \nu}F^{\mu \nu}$ vanishes
\cite{SD}. Taking into account the expression(\ref{Bbar}) for the reduced
magnetic field, this assumption reads, in conformal coordinates,
\be \lb{FF}
\frac{1}{2}F_{\mu \nu}F^{\mu \nu} = A_0^2\,\left[ \,\frac{4e^4}{\mu^2}(\phi\st
\phi)^2 - {\rm e}^{-2u}\,\left( \frac{\nabla A_0}{A_0} \right)^2\, \right]
= 0\,.
\ee
In the one-vortex case, we chose polar coordinates $\ol{r}$, $\theta$ such
that
\ba \lb{stat rot}
& ds^2 = & (\,dt + Y(\ol{r})\,d\theta)^2 - \sigma^2(\ol{r})\,d\theta^2 -
d\ol{r}^2\,, \nonumber\\
& \phi = & R(\ol{r}) {\rm e}^{in\theta}
\ea
(the proper radial distance $\ol{r}$ is related to the conformal
radial distance $r$ defined from equation (\ref{conf}) by $d\ol{r} = {\rm
e}^u\,dr$). The square root of the ansatz (\ref{FF}) then leads to the
differential equation
\be \lb{FF2}
A_0^{-1}\,\frac{dA_0}{d\ol{r}} = \varepsilon\, \frac{2e^2}{\mu}\,
\phi\st \phi\,,
\ee
which (taking into account equation (\ref{A_0})) is solved by
\be \lb{R2}
R^2 = \frac{\eta^2}{1+{\rm e}^{\ds \,- \frac{2e^2 \eta^2}{|\mu|}\,
(\ol{r} - \ol{r}_0)}}\,,
\ee
where $\ol{r}_0$ is an integration constant, and we have chosen without
loss of generality $\varepsilon = - {\rm sign}(\mu)$. Clearly, the variable
$\ol{r}$ is unbounded; as $\ol{r}$ varies over the real axis, the Higgs
field interpolates monotonously between the symmetrical vacuum $R = 0$ and
the asymmetrical vacuum $R = \eta$, and may be chosen as new radial
coordinate. We then find that, for any value of the parameter $a$, equation
(\ref{h}), in which we choose $h(z) = {\rm const.} \, z^{1+k}$ ($k$
constant), is consistent with the two coupled non-linear Schr\"odinger
equations for $R$ and $\omega$ mentioned above, which may be solved to
yield the explicit metric
\be \lb{sol}
ds^2 = (\,dt + \varepsilon (\sigma - c)\,d\theta)^2 - \sigma^2\,d\theta^2
- \frac{\mu^2}{e^4}\,\frac{dR^2}{R^2(\eta^2 - R^2)^2} \,,
\ee
with
\be \lb{sig}
\sigma = p\,R^{\ds 8\pi G\eta^2(1 - \frac{\mu^2a^2}{e^2\eta^4})}\,
(\eta^2 - R^2)^{\ds \frac{4\pi G \mu^2 a^2}{e^2\eta^2}}\,
{\rm e}^{\ds -4\pi G R^2}
\ee
($p\,$ positive constant), and $c = - \varepsilon k/8\pi G \mu a^2$. The
associated electric potential $A_0$ is given by equation (\ref{A_0}), while
the magnetic potential $A_{\theta} = \ol{A}_{\theta} + Y A_0$ is determined
from equation (\ref{S3}) to be
\be \lb{sol2}
A_{\theta} = - \varepsilon c A_0 - \frac{n}{e} \,.
\ee
The metric (\ref{sol}) and the gauge potentials (\ref{sol2}) are regular
over the range of variation of the Higgs field R.

For a metric of the form (\ref{sol}), $b = \varepsilon \,d\ln \sigma/d\ol{r}$,
leading to the spatial curvature
$R_{\gamma} = 2(b^2 + \varepsilon db/d\ol{r})$.
At either endpoint $R = \eta$ ($\ol{r} = + \infty$) or $R = 0$
($\ol{r} = - \infty$), the twist $b$ goes to a constant, while its first
derivative vanishes from equation (\ref{FF2}). We then find from
equation (\ref{Ricci}) the asymptotic behaviours of the Ricci tensor
\be \lb{Lambda}
R_{\mu \nu} \simeq \frac{1}{2} b^2({\scriptstyle \pm \infty}) \, g_{\mu \nu}
\;\;\;\; (\ol{r} \ra \pm \infty) \,,
\ee
so that the regular spacetime (\ref{sol}) is, for any value of $a$,
asymptotic to two spacetimes of
constant curvature corresponding to negative cosmological constants
$\Lambda_{\pm} = -l_{\pm}^{-2}$, with $l_{\pm}^{-1} = b({\scriptstyle
\pm \infty})/2$ (comparing (\ref{Lambda}) with the first equation
(\ref{eqs}), we check that this effective cosmological constant is due to
the potential energy $V({\scriptstyle  \pm \infty}) = - b^2({\scriptstyle
\pm \infty})/32 \pi G$). More precisely, the metric (\ref{sol}) goes over
to
\be \lb{as}
ds^2 \simeq (\,dt + \varepsilon (\sigma - c) \,d\theta)^2 - \sigma^2
\,d\theta^2 - \frac{l_{\pm}^2}{4} \,\frac{d\sigma^2}{\sigma^2} \;\;\;\;
(\ol{r} \ra \pm \infty) \,.
\ee
The asymptotic metric (\ref{as}) is the extreme BTZ solution \cite{BTZ}
with negative mass $M = \varepsilon l_{\pm}^{-1} J = -2c^2/l_{\pm}^2$,
viewed in an uniformly rotating frame with angular velocity $\Omega =
\varepsilon l_{\pm}^{-1}$ and with time rescaled by $t \ra (l_{\pm}/2c)\,t$;
the apparent singularity at $\sigma = 0$ is actually at infinite geodesic
distance \cite{EL}. The asymptotic behaviours of $\sigma$ are, for $G >
0$, $\sigma(+ \infty) = + \infty$ for $a^2 < 0$, $\sigma(+ \infty) = 0$
for $a^2 > 0$, and $\sigma(- \infty) = 0$ for $a^2 < (e^2/\mu^2)\eta^4$,
$\sigma(- \infty) = + \infty$ for $a^2 > (e^2/\mu^2)\eta^4$. For $G < 0$ (the
sign of the gravitational constant is not fixed {\em a priori} in three
dimensions \cite{DJH}), these asymptotic behaviours are inversed, with
$\sigma(\pm \infty)$ replaced by $1/\sigma(\pm \infty)$. For the
particular values $a^2 = 0$, or $a^2 = (e^2/\mu^2)\eta^4$, the metric
(\ref{sol}) is asymptotic for $\ol{r} \ra + \infty$ (resp. $- \infty$), to
the flat metric
\be \lb{as0}
ds_0^2 = [\,dt + \varepsilon (\sigma_{\infty} - c)\,d\theta \,]^2 -
\sigma_{\infty}^2 \,d\theta^2 - d\ol{r}^2
\ee
($\sigma_{\infty}$ constant), which is simply the cylindrical Minkowski
metric $ds^2 = dt^2 - \sigma_{\infty}^2 \,d\theta^2 - d\ol{r}^2$, viewed in
an uniformly rotating frame and time rescaled; the doubly self-dual
solution (26)-(29) thus automatically satisfy, for these values of $a^2$,
the flat-space boundary conditions (20) at one of its two endpoints.
For any value of $a$, our solution interpolates between two inequivalent
vacuum configurations, and is thus a topological soliton. However,
contrary to the case of vortex solutions with Euclidean spatial topology,
the magnetic flux
\be \lb{flux}
\Phi = A_{\theta}(+ \infty) - A_{\theta}(- \infty) = \mp c \frac{e}{|\mu|}
\eta^2
\ee
is unquantized for our solution. The reason is that the usual flux
quantization follows from the condition that the vector potential is
regular at the origin of polar coordinates; this condition does not hold
in our cylindrical topology (the {\em apparent} flux at $+ \infty$,
$\Phi_+ \equiv A_{\theta}(+ \infty)$, is from equations (\ref{sol2}) and
(\ref{bc1}) quantized for $a = 0$).

Our solution (\ref{R2}), (\ref{A_0}), (\ref{sol2}) for the Higgs and gauge
fields does not depend on the gravitational constant $G$, and thus
survives in the limit $G \ra 0$. If we also take in this limit $a^2 \ra
\infty$ so that $|Ga^2| \equiv 1/4\pi|\mu|l$ stays fixed, we obtain the
exact vortex solution (\ref{R2}), (\ref{A_0}), (\ref{sol2}) sitting on the
background constant curvature metric (\ref{as}), which in the case $a^2$
finite ($l \ra \infty$), reduces to the flat cylindrical metric (\ref{as0}).
This last solution was previously obtained by Schiff \cite{Schiff} (the
squared Higgs field is given in equation (2.20) of \cite{Schiff}, and the
background metric in equation (2.29)), who pointed out that the
Minkowskian flux quantization law is lost. Let us also recall that Comtet
and Khare \cite{C+K} showed that the Maxwell-Chern-Simons-Higgs model with
the standard fourth-order potential admits self-dual vortex solutions on a
specific background metric with constant curvature, and $g_{00} = 0$
(the constant $k_1$ in equation (11b) of \cite{C+K} may be taken equal to
zero without loss of generality); probably much of the analysis of the
present paper could also be carried out starting from the assumption
$g_{00} = 0$, instead of (\ref{ans1}).

An unpleasant feature of the metric (\ref{sol}) is the occurrence of
closed timelike curves (CTCs) for $c\, \sigma(R) < c^2/2$. All the circles
$t = $ const., $R = $ const. are CTCs for $c < 0$, and CLCs (closed
timelike lines) for $c = 0$. For $c > 0$, CTCs occur whenever $\sigma(R)$
goes to 0 at one endpoint. So the only CTC-free case is $c > 0$, $G <0$,
$0 \leq a^2 \leq (e^2/\mu^2)\eta^4$; then $\sigma$ goes to $+ \infty$
(or a positive constant value) at both endpoints and no CTCs occur
provided the constant $p\,$ in (\ref{sig}) is large enough. Another
advantage of the choice $G < 0$ is that the potential (\ref{V}), which is
unbounded from below for $G > 0$ \cite{C+L}, is now positive definite, with
the extrema $R = 0$ and $R = \eta$ remaining local minima if $|G|$ is not
too large\footnote{Menotti and Seminara have shown \cite{M+S} that if the
weak energy condition is satisfied, there are no CTCs in (2+1) dimensions.
The occurrence of CTCs in our solution for $G > 0$ is thus due to the fact
that the potential is not positive definite. There is no obvious reason
for the existence of CTC-free solutions for $G < 0$, as the weak energy
condition for the source $8\pi G T_{\mu \nu}$ is certainly not satisfied in
this case ($G T_{00}$ is negative definite).}.

We have shown that the sole assumption $g_{00} = 1$ leads to self-dual
stationary solutions of the Chern-Simons-Higgs model (with a suitable
eighth order potential) coupled to Einstein gravity, and derived a set of
exact, doubly self-dual solutions. These solutions interpolate between the
two vacua (symmetrical and asymmetrical), the corresponding spacetimes
being asymptotic to two spacetimes of constant curvature. The next,
logical step would be to investigate the existence of self-dual vortex
solutions to the Chern-Simons-Higgs model coupled to topologically massive
gravity \cite{DJT}, or even to pure Chern-Simons gravity \cite{CSS}.

\vskip 1cm

\noindent {\bf Acknowledgment}

I wish to thank Peter Horv\'athy for the kind invitation to the Tours
Workshop on Chern-Simons gauge theories (11-12 Sept.\ 1995), which prompted
this research.

\end{document}